\documentclass[aps, prl, floatfix, superscriptaddress, showpacs, 12pt]{revtex4-1}

\usepackage{graphicx}
\usepackage{color}
\usepackage{calc}
\usepackage{array}
\usepackage{graphicx}
\usepackage{amsmath, amssymb}
\usepackage{natbib}
\usepackage{hyperref}

\DeclareMathOperator{\dd}{\mathrm{d}\!}

\begin{document}

%% ----------------------------------------------------------------------------
%%
%% TITLE PAGE
%%

\title{First-principles approach to thin superconducting slabs and heterostructures}

\author{G{\'{a}}bor \surname{Csire}}%
\affiliation{Institute for Solid State Physics and Optics,
             Wigner Research Centre for Physics, Hungarian Academy of Sciences, \\
             PO Box 49, H-1525 Budapest, Hungary}

\author{Stephan Sch{\" o}necker}
\affiliation{Applied Materials Physics, Department of Materials Science and Engineering,\\
              Royal Institute of Technology, Stockholm SE-10044, Sweden}

\author{Bal{\'a}zs {\'U}jfalussy}
\affiliation{Institute for Solid State Physics and Optics,
             Wigner Research Centre for Physics, Hungarian Academy of Sciences, \\
             PO Box 49, H-1525 Budapest, Hungary}

\date{\today}

\begin{abstract}

We present a fully first-principles method for superconducting thin films.
The layer dependent phonon spectrum is calculated to determine
the layer dependence of the electron-phonon coupling for such systems,
which is coupled to the Kohn-Sham-Bogoliubov-de Gennes equations, and it is solved in a parameter free way. The theory is then applied to different surface facets of niobium slabs and to niobium-gold heterostructures. We investigate the dependence of the transition temperature
on the thickness of the slabs and the inverse proximity effect observed in thin superconducting heterostructures.

\end{abstract}

%\pacs{63.20.kd, 74.25.Jb, 74.45.+c, 74.78.Fk}
% PACS numbers:
% 63.20.kd	Phonon-electron interactions
% Surface states, 73.20.-r
% electronic structure calculations in superconductivity, 74.20.Pq
% Andreev reflection (superconductivity), proximity effects, SN and SNS junctions, 74.45.+c
% electronic structure properties of superconductors, 74.25.Jb
% superconducting heterostructures, 74.78.Fk
% Ab initio calculations (electronic structure of atoms and molecules), 31.15.A-
% \keywords{}

\maketitle

Thin film superconductivity is a subject of great scientific interest
since the 1950s~\cite{Buckel1954,Blatt1963,Abeles1966,Strongin,Haviland1989}. 
The development of nanotechnology has led to the renaissance of this
topic~\cite{Bourgeois,Guo2004,zer2006,Eom2006,Brun2009,Wang2009,Qin2009,Zhang}
due to possible technological applications in superconducting
nanodevices.
Theoretically, it is entirely possible for thin (few nanometers thick) slabs
that a large electron-phonon coupling at the surface can lead to
superconductivity well above the bulk transition temperature.
For such superconducting heterostructures an inverse proximity effect was observed
in Ref.~\cite{Bourgeois}:  a non-superconducting metal overlayer on
a superconducting thin film increases the critical temperature $T_c$.
This is in strong contrast to the case of the thick (compared to the coherence length) superconducting films, 
where the metallic overlayer decreases $T_c$~\cite{Yamazaki1,Yamazaki2,tcpaper}. 
In this paper, the main focus is how the material specific, intrinsic superconducting properties
(which are essential for technological applications) change as a function of the thickness.
In the case of thin superconducting layers the electron-phonon interaction may change significantly,
which can lead to new and interesting effects.
To properly describe such a situation, a fully first-principles approach is needed,
which takes into account the changes in the electronic structure 
and in the phonon spectrum.  
However, the simultaneous treatment of vibrational and electronic degrees of freedom on the same level
leads to complications which are very difficult to overcome. 
Here we propose a simplified treatment, where both spectrums are calculated on the first-principles level separately and the results are combined.

The density functional theory (DFT) for superconductors yields the
Kohn-Sham-Bogoliubov-de~Gennes (KSBdG) equations~\cite{OGK,Luders1,Luders2}
by introducing the $\chi(\vec r)=\left< \Psi_\downarrow(\vec r) \Psi_\uparrow(\vec r) \right>$ anomalous density
as an additional density, analogously to the magnetization in spin-polarized DFT theory.
In the case of multilayer systems, the self-consistent solution of these equations can be obtained in terms of the 
Screened Korringa-Kohn-Rostoker (SKKR) method (see Ref.~\cite{BdGKKR}),
where the retarded Green-function,  $\{G^{ab,+}_{IJ,LL'}(\varepsilon, \vec r,\vec{k}_{||}) \}$
is the fundamental quantity of interest. Here $a,b$ refer to the electron-hole components, $I,J$ are the layer indices 
and  $L=(l,m)$ is a composite angular momentum index.
Physical quantities, like the $\rho_I(\vec r)$ charge and $\chi_I(\vec r)$ anomalous densities
can be calculated from the layer diagonal Green-function as it was described in Ref.~\cite{BdGKKR}.
For self-consistent calculations we use the parametrization for the exchange energy introduced by Suvasini et al.~\cite{Suvasini}
\begin{equation}
 E_{xc,I}[\rho_I,\chi_I] = E^0_{xc,I}[\rho_I] - \int  \chi_I^*(\vec r) \Lambda_I \chi_I(\vec{r})  \dd \vec{r},
 \label{eq:exchange}
\end{equation}
where $E^0_{xc,I}[\rho_I]$ is the usual exchange correlation energy for electrons in the normal state
and $\Lambda_I$ describes the strength of the electron-phonon interaction for layer $I$.
Each layer is assumed to be chemically homogeneous, but any two distinct layers can, in principle,
describe different materials constituents.

The approximation (\ref{eq:exchange}) to the exchange-correlation potential
takes into account the electron-phonon interaction via a single layer dependent parameter. This parameter 
can be estimated from the $\lambda_I$ electron-phonon coupling constant as
$\Lambda_I=\lambda_I/D_I(E_F)$, where $D_I(E_F)$ is the density of states (DOS) at the Fermi-energy for layer $I$.
Furthermore, the electron-phonon coupling constant can be calculated as~\cite{McMillan}:
\begin{equation}
  \lambda_I=\frac{D_I(E_F) \left< g^2_I \right>}{M_I \left< \omega_I^2 \right>},
  \label{eq:McMillan1}
\end{equation}
where $M_I$ is the atomic mass, and $D_I(E_F) \left< g_I^2 \right>$ is the McMillan-Hopfield parameter.
One can immediately recall that various theories~\cite{Gaspari,Savrasov,Savrasov2} have been worked out 
in the literature to calculate the terms in the above expression. A 
purely electronic calculation leads to the McMillan-Hopfield parameter via the Gaspari-Gy{\H o}rffy formula~\cite{Gaspari},
which is based on the following assumptions:
(i) the atomic potentials are spherically symmetric,
(ii) neglects every special influence of the shape of the Fermi surface,
(iii) small displacements in the atomic potential can be approximated by a rigid shift. 
The other important parameter is the average of the square of the phonon frequency, 
 $\left< \omega_I^2 \right>$, and can be calculated based on the formula~\cite{McMillan}:
\begin{equation}
  \left< \omega_I^2 \right> \approx \frac{\int \dd \omega ~\omega~ F_I(\omega)}{\int \dd \omega ~\frac{1}{\omega}~ F_I(\omega)},
\end{equation}
where $F_I(\omega)$ is the phonon DOS for layer $I$.
Even at this point one can notice that a larger McMillan-Hopfield parameter or 
the softening of $\left< \omega_I^2 \right>$ will result in a larger electron-phonon coupling.

Our phonon calculations are based on relaxed slab geometries, and interlayer relaxations are assumed
for all interlayer distances perpendicular to the surface facets with fixed in-plane lattice parameter.
The first principles calculation of dynamical properties of lattices requires the knowledge of interatomic forces.
We determine the force constant matrix for bulk, slabs, and heterostructures
in the framework of density functional perturbation theory~\cite{Baroni2001}
as implemented in the Vienna ab-initio simulation package (VASP)~\cite{Kresse1996}
and employing Phonopy~\cite{Togo2015} to compute the dynamical matrix and layer resolved phonon DOSs. 
Once the layer dependent phonon spectrum has been obtained, the layer dependent electron-phonon coupling
constants can be calculated based on Eq.~(\ref{eq:McMillan1}) and, consequently, the KSBdG equations
can be solved self-consistently for finite temperatures with the SKKR method. 
These self-consistent calculations are carried out within the atomic sphere approximation with an
angular momentum cutoff of $l_{max} = 2$.
In order to determine the superconducting transition temperature, 
one needs to find the critical temperature where the spectrum of the KSBdG Hamiltonian does not give a gap.

In what follows, we choose niobium as the testbed and primary target of our numerical investigations.
To verify the theory, we first calculated the electron-phonon coupling and the critical temperature for bulk Nb,
and obtained $\lambda=0.86$ and $T_c=11.3$~K.
Based on the Gaspari-Gy{\H o}rffy theory using the augmented plane wave method for Nb $\lambda=0.88$
was obtained by Klein and Papaconstantopoulos~\cite{Klein}.
In Ref.~\cite{Luders2} a multicomponent DFT for the combined system of electrons and nuclei
with different hybrid functionals led to critical temperatures in the range of 8.4 - 9.5~K,
while the known experimental bulk values for 
Nb are~\cite{Klein}: $\lambda^{(exp)}=0.82$ and $T^{(exp)}_c=9.2$~K.
It can be seen that our results are rather similar to the results of other authors for the electron-phonon
interaction, and slightly overestimated the critical temperature compared to experiments, which, despite
the simplicity of the used exchange energy, still not far from the experimental value.
Here it is worth mentioning that in the case of niobium, phonon retardation effects
play an important role, therefore it should be treated in the strong coupling limit.
In our theory the anomalous density $\chi_I(\vec r)$ influences the effective potential
$V_{eff,I}(\vec r)$ via the density $\rho_I(\vec r)$, which is the analogy of the self-energy correction
to the Eliashberg equations~\cite{Allen1983} and may be regarded as a strong coupling effect.

Now we are ready to apply the method to niobium slabs, and niobium -- gold heterostructures. 
It should be noted, that throughout the whole paper we neglect the effect of a substrate which could, in principle,
modify the results quantitatively, but should not alter the basic physics
and would just lead to numerical complications in the calculation of the phonon spectrum. 
In the case of a Nb slab, the calculations were performed for 3,6,9,12 and 15 layers of Nb.
We choose two facets for our studies, the open (100) surface facet, because it is the most stable surface facet, and
a contrasting close-packed one, namely the (110) facet which has a slightly higher surface energy.

\begin{figure}[hbt!]
   \includegraphics[width=0.5\linewidth]{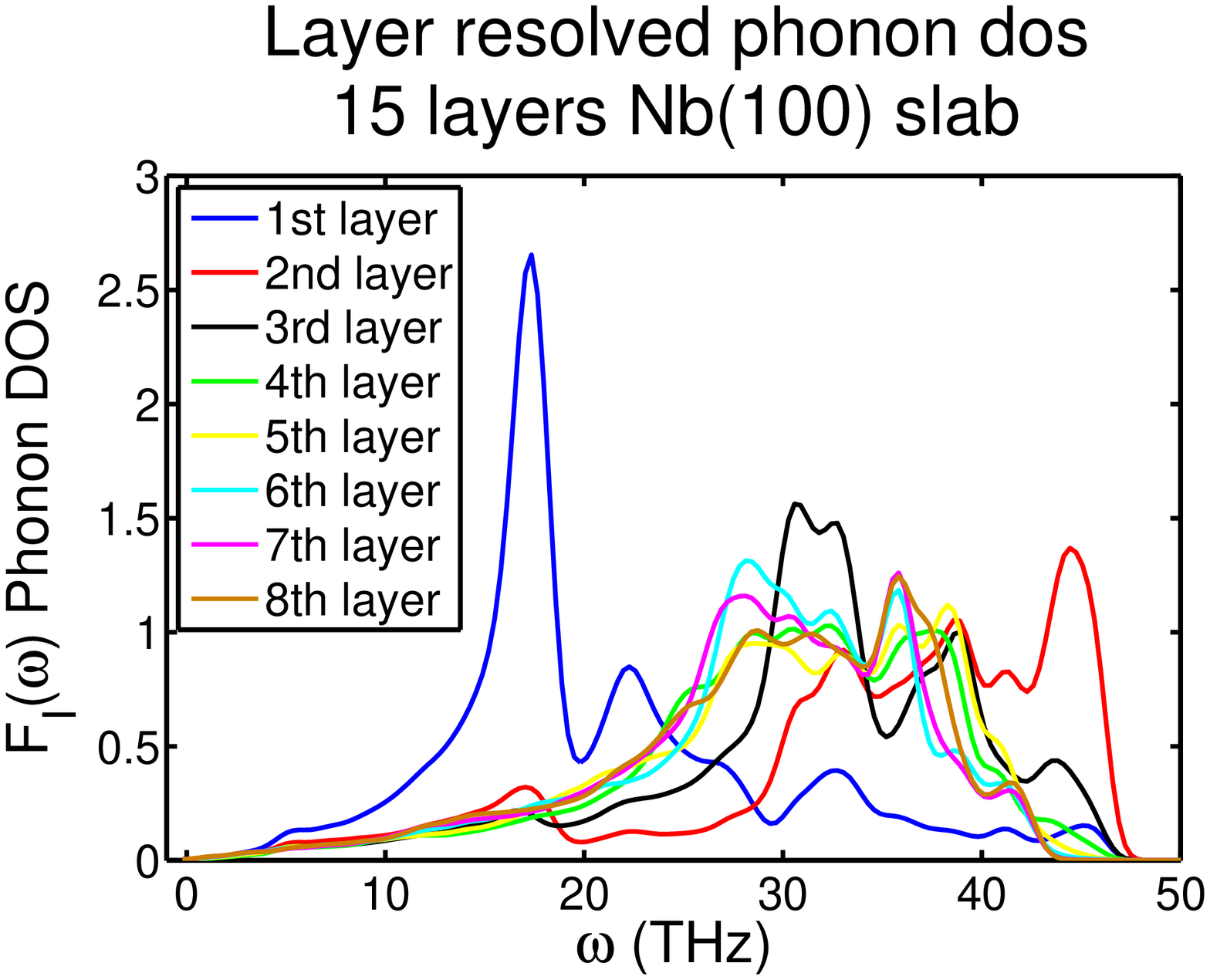}~
   \includegraphics[width=0.5\linewidth]{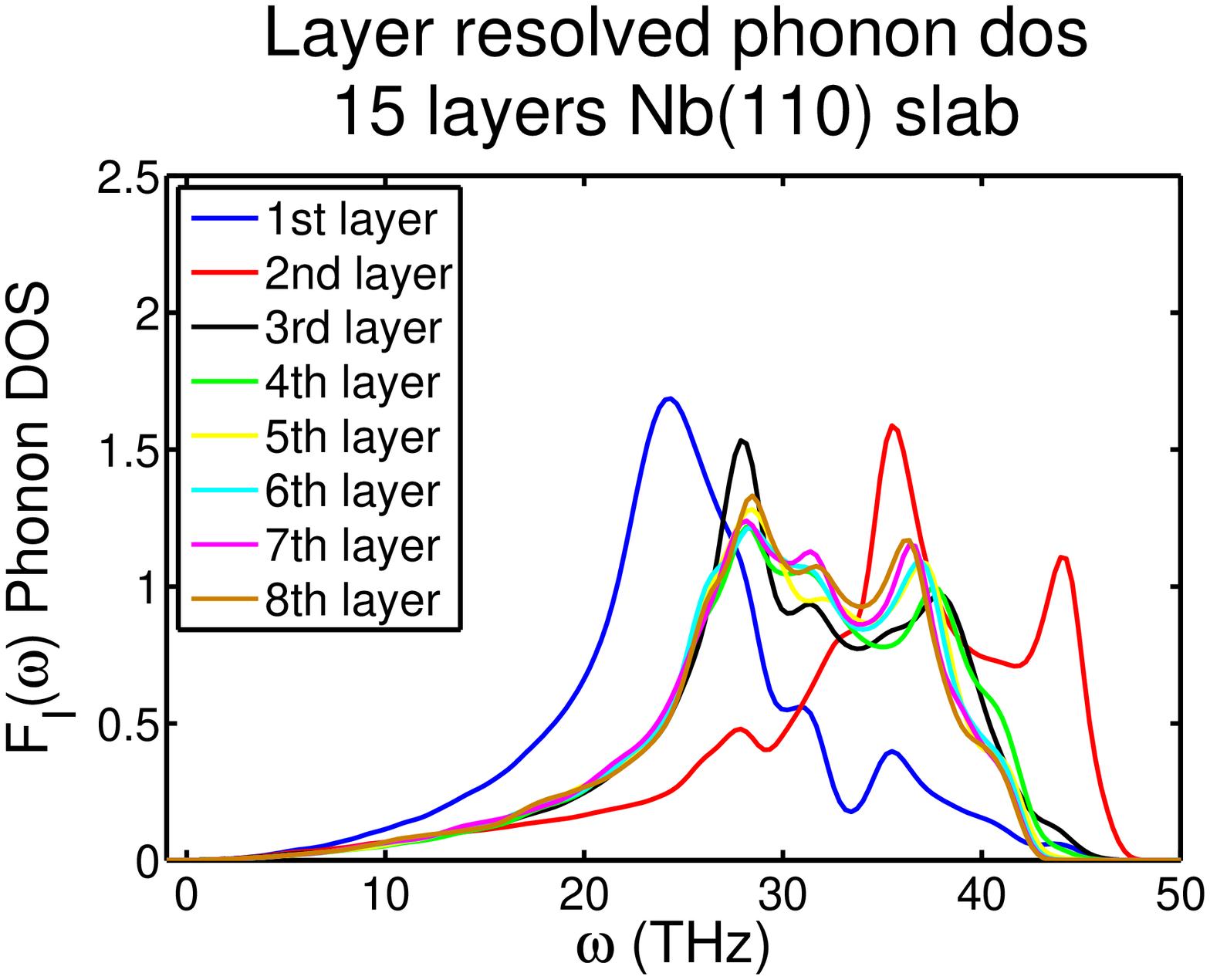}
   \caption{\label{fig:nb15_eph}%
            (Color online) Layer resolved phonon DOS of 15 layers Nb slab 
            for bcc(100) (left panel) and bcc(110) (right panel).
            The first layer is the surface layer, second layer is the subsurface layer, ... 8th layer is the bulk-like center of the slab.}
\end{figure}

\begin{figure}[hbt!]
 \centering
   \includegraphics[width=0.5\linewidth]{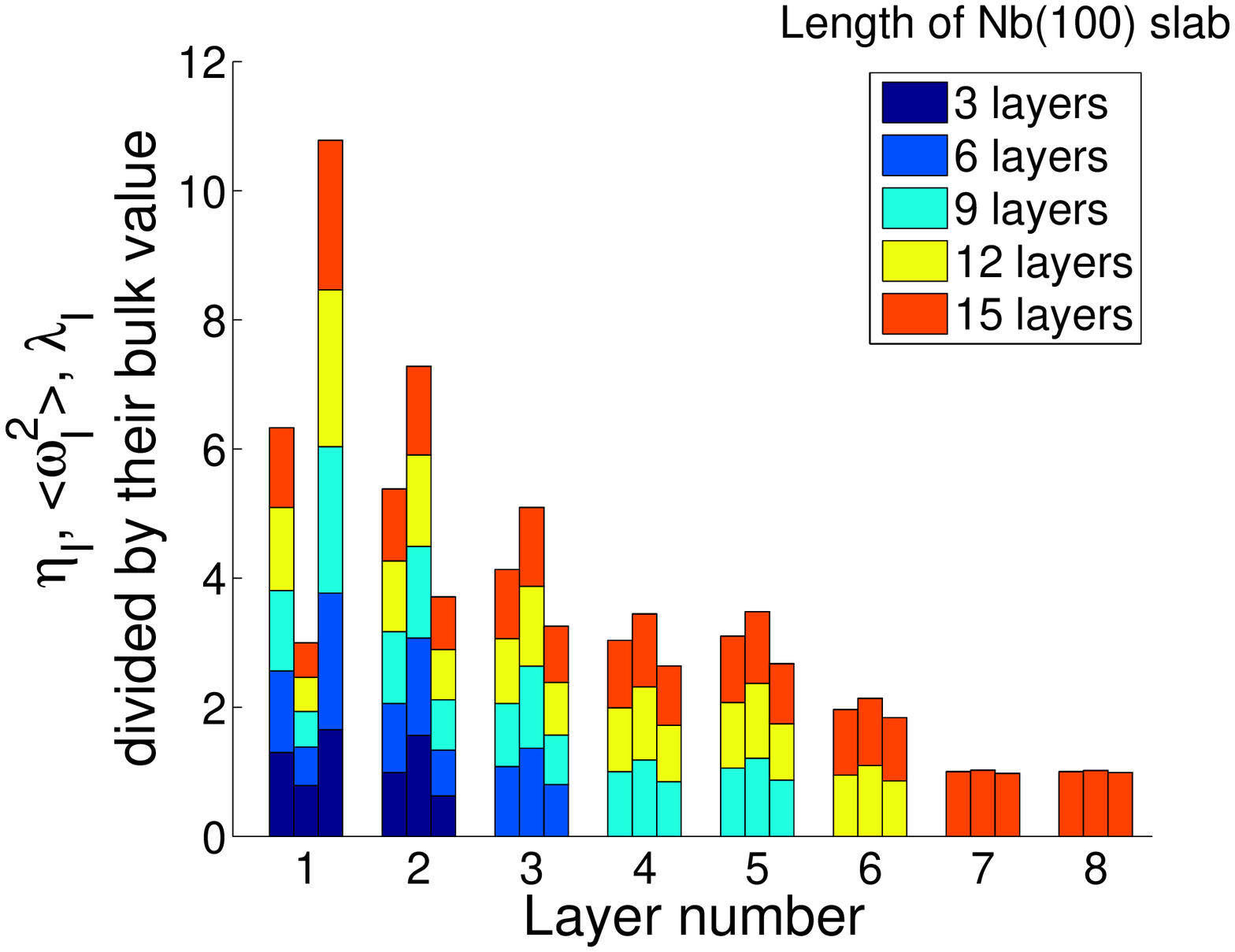}\\
   \includegraphics[width=0.5\linewidth]{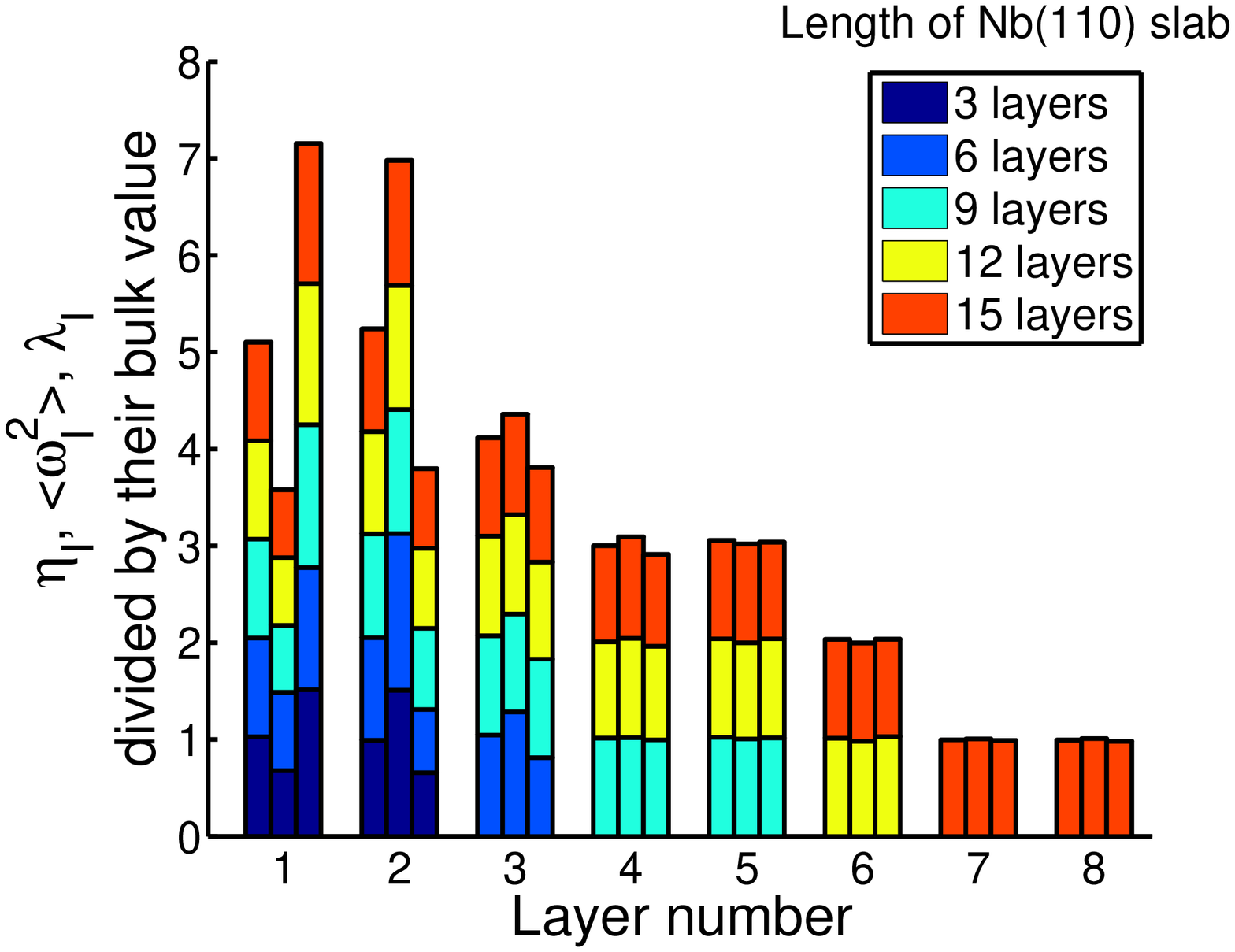}\\
   \includegraphics[width=0.5\linewidth]{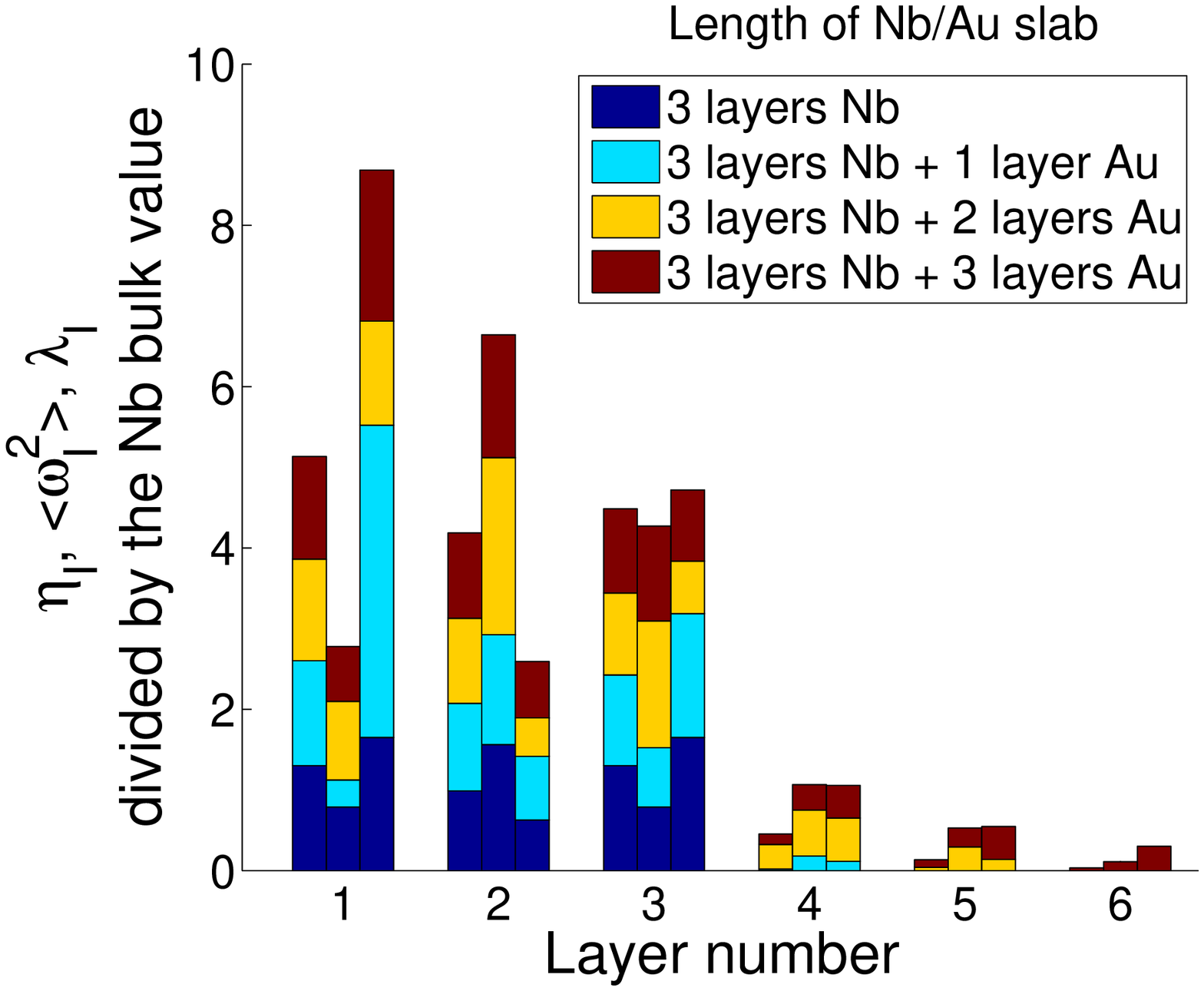}
   \caption{\label{fig:sum_nb}%
          (Color online) $\eta_I=D_I(E_F) \left< g^2_I \right>/M_I$, $\left< \omega_I^2 \right>$, $\lambda_I$ -- normalized with
          the Nb bulk value -- are shown in the stacked bar charts
          (the actual value is always added to the sum of the other data sets), respectively in each bar, for different lengths
          of Nb(100) (top panel), Nb(110) (middle panel) and Nb/Au (bottom panel) slabs, where 1,2 and 3 layers of Au were added on 3 Nb layers.}
\end{figure}

The results obtained for the McMillan-Hopfield parameter, the average phonon frequency and the electron-phonon interaction
are presented in a graphical form with a stacked bar chart in Fig.~\ref{fig:sum_nb}.
Since the slabs are symmetric with respect to the center of the sample,
we plot the results only from the surface layer to the middle of the sample.
In Fig. \ref{fig:nb15_eph} the phonon DOS is shown
for both Nb(100) and Nb(110) slabs consisting of 15 atomic layers.
It can be observed that as approaching the middle of the sample the phonon DOS converges.
Faster convergence was obtained in the case of the electron DOS (not shown).
It can be seen that on the surface of the bcc(100) slab,
the phonon DOS is dominated by low frequency states, therefore, 
the $\left< \omega_I^2 \right>$ becomes significantly smaller just on the first surface layer
(see Fig.~\ref{fig:sum_nb}). This effect can also be observed for the bcc(110) slab
but it is not as pronounced, and mostly compensated by the subsurface layer where the phonon DOS is dominated by high frequency states.
As a consequence, for the bcc(100) surface facet, the McMillan-Hopfield parameter increases on approaching the surface, which
is in sharp contrast to the bcc(110) surface facet where the McMillan-Hopfield parameter fluctuate around its bulk value for all layers.
At the (100) surface, both the electron and the phonon parts
enlarge the electron-phonon coupling significantly beyond the bulk value.
At the subsurface layer the electron-phonon coupling becomes smaller because of the larger $\left< \omega_I^2 \right>$,
and as we approach the middle of the sample its value converges to the bulk value.
In the case of bcc(110) slab the electron-phonon coupling changes similar to the bcc(100), 
however, an important difference is that on the surface the electron-phonon coupling is not as large. %% as in the case of a bcc(100) slab.

Once knowing the electron-phonon interaction parameters for all layers,
we can proceed and solve the KSBdG equations self-consistently for various temperatures.
In the case of $T=0$~K we find that the superconducting gap has a layer dependence (not shown), which 
follows the layer dependence of the $\lambda_I$ electron-phonon coupling parameter.
However, when the KSBdG equations are solved for finite temperatures, it is found that in all layers the superconducting gaps
disappear at the same critical temperature.
This means that a layer, which has a larger electron-phonon coupling parameter,
strengthens the superconducting properties of the other layers
with smaller electron-phonon coupling via the proximity effect~\cite{Proximity}.
Formally, this is very similar to the case of MgB$_2$'s two bands system~\cite{Nicol2005} where
the two superconducting gaps have the same critical temperature only if there is an interband coupling.

\begin{figure}[hbt!]
 \centering
   \includegraphics[width=1\linewidth]{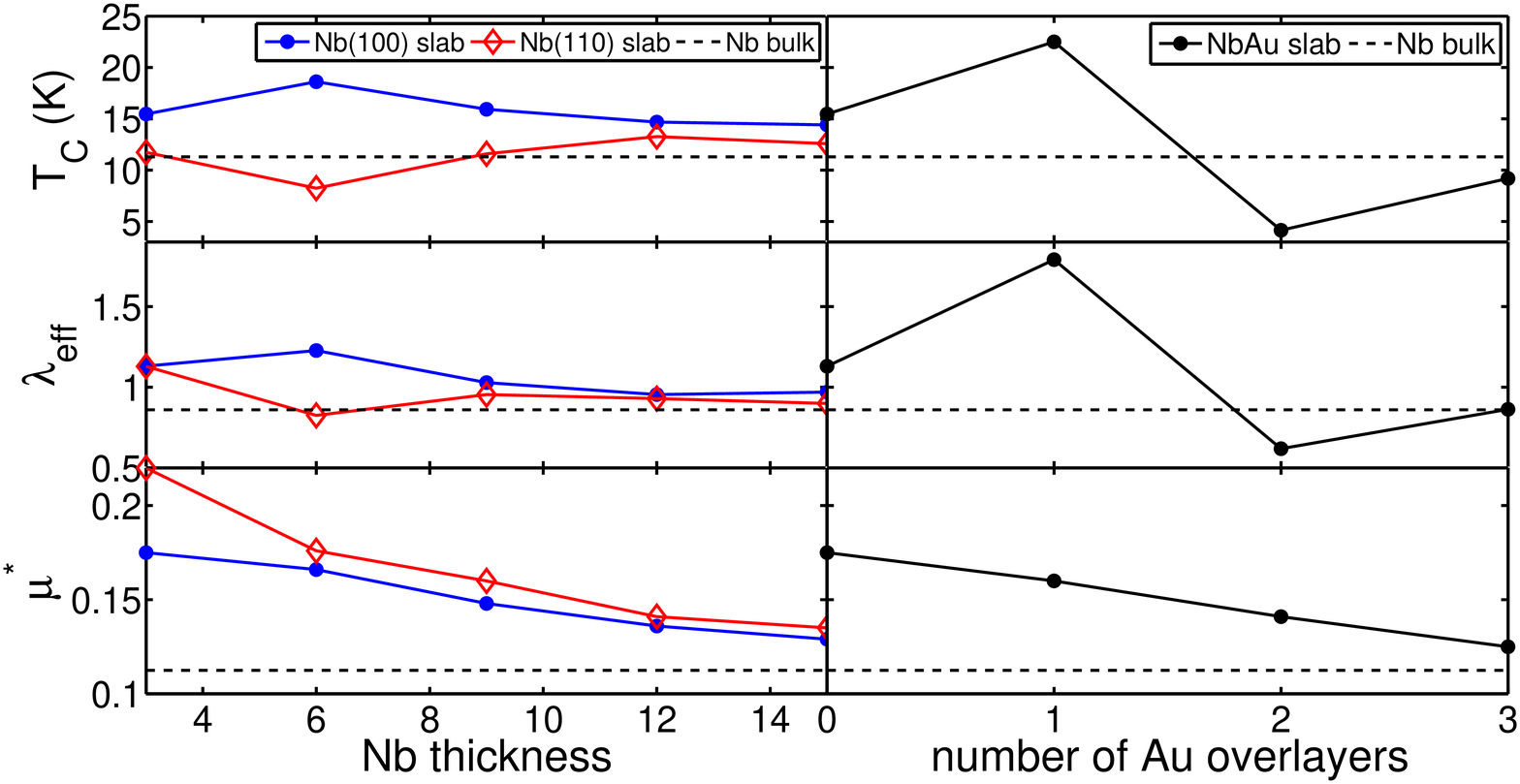}
   \caption{\label{fig:mcm_nb}%
          (Color online) The critical temperature ($T_c$), effective electron-phonon coupling ($\lambda_{eff}$)
           and Coulomb repulsion ($\mu^*$) as function
          of the thickness of Nb(100) (blue line, full symbol) and Nb(110) (red line, open symbol) slab (left panels) and Nb/Au slabs (right panels).}
\end{figure}

In Fig.~\ref{fig:mcm_nb} (top left panel) it can be seen that the critical temperature of the Nb(100) slab
is well above the bulk critical temperature (with a maximum at the 6 layers thick Nb slab),
which is clearly due to the larger electron-phonon coupling on the surface. 
Not surprisingly, the Nb(110) slab's critical temperature is always lower than in the case of Nb(100) slab. 
In order to gain deeper understanding of the changes in the critical temperature due to the thickness, it is interesting
to look at  other properties of superconducting slabs, such as $\mu^*$, the effective Coulomb repulsion.
The $\mu^*$ is a fundamental quantity in the theory of superconductivity, related to the
correlation effects due to the Coulomb repulsion.
Usually, it is treated as an adjustable parameter, but based on the previous results, it is possible to estimate it thin film systems.
For a strong-coupling superconductor like Nb, $T_c$ is given by the McMillan formula~\cite{McMillan},
which depends on the Debye temperature $\Theta_D$,
the effective electron-phonon coupling $\lambda_{eff}$ and the $\mu^*$.
If the values of $\lambda_{I}$  are known, it is possible to calculate $\lambda_{eff}$ as~\cite{deGennes}:
\begin{equation}
\lambda_{eff}= \frac{ \sum_I \lambda_{I} D_I(E_F) }{\sum_I D_I(E_F)},
\end{equation}
where $I$ is a layer index.  
$\Theta_D$ can be obtained from the phonon spectrum.
Thus the $\mu^*$ can also be calculated by equating the value of $T_c$ obtained previously to the McMillan formula. 
The results are shown in Fig.~\ref{fig:mcm_nb} (left panels), where one can see that 
the effective Coulomb repulsion is decreasing as a function of the niobium thickness.
%% (more rapidly for the bcc(110) structure).
This is probably due to the fact that for thicker slabs the electrons have more degrees of freedom.
It is also worth mentioning that as it can be seen in Fig.~\ref{fig:mcm_nb}, the superconducting transition temperature has a 
rather similar  dependence on
the thickness of the slab as the above defined $\lambda_{eff}$.

\begin{figure}[hbt!]
   \includegraphics[width=0.5\linewidth]{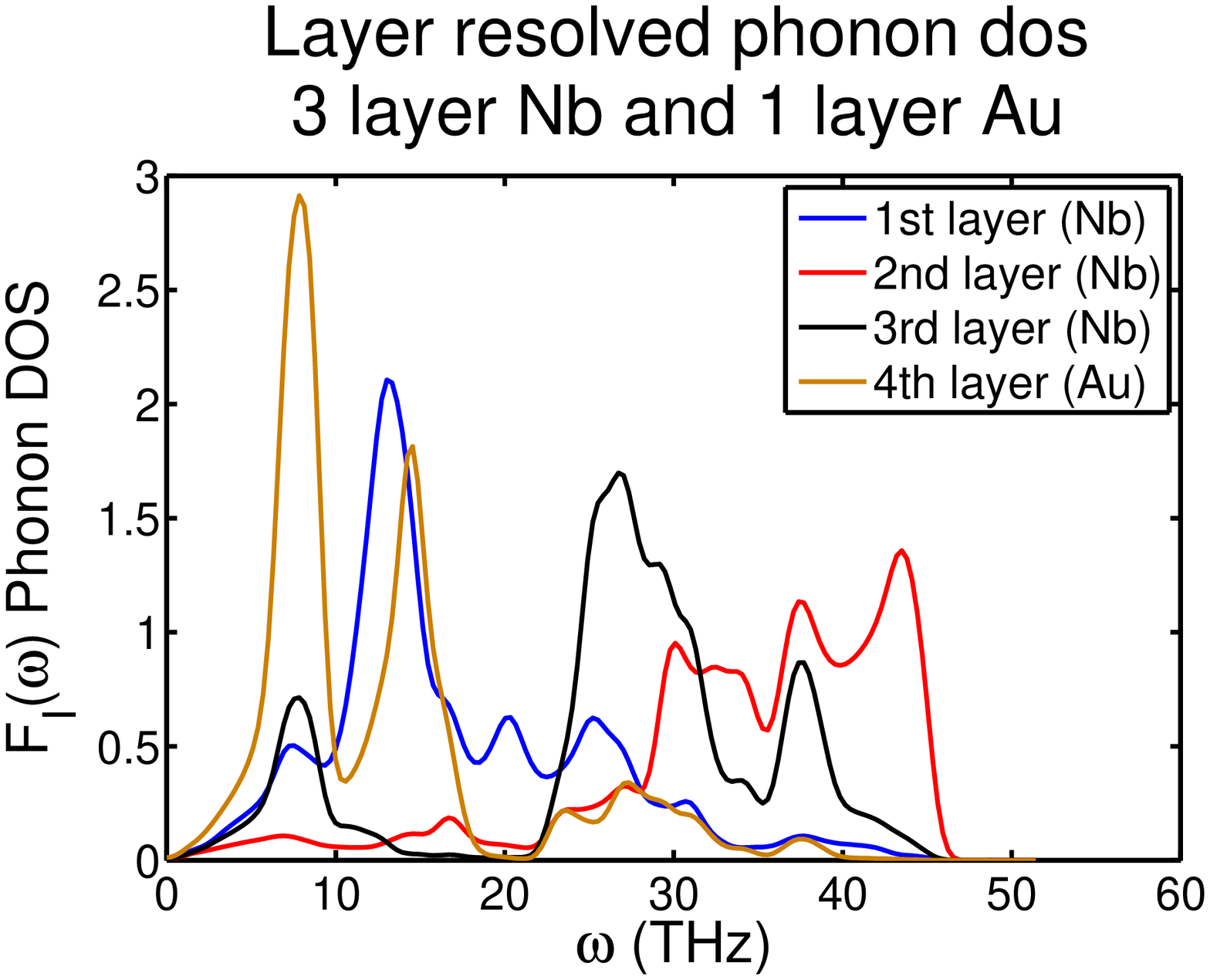}~
   \includegraphics[width=0.5\linewidth]{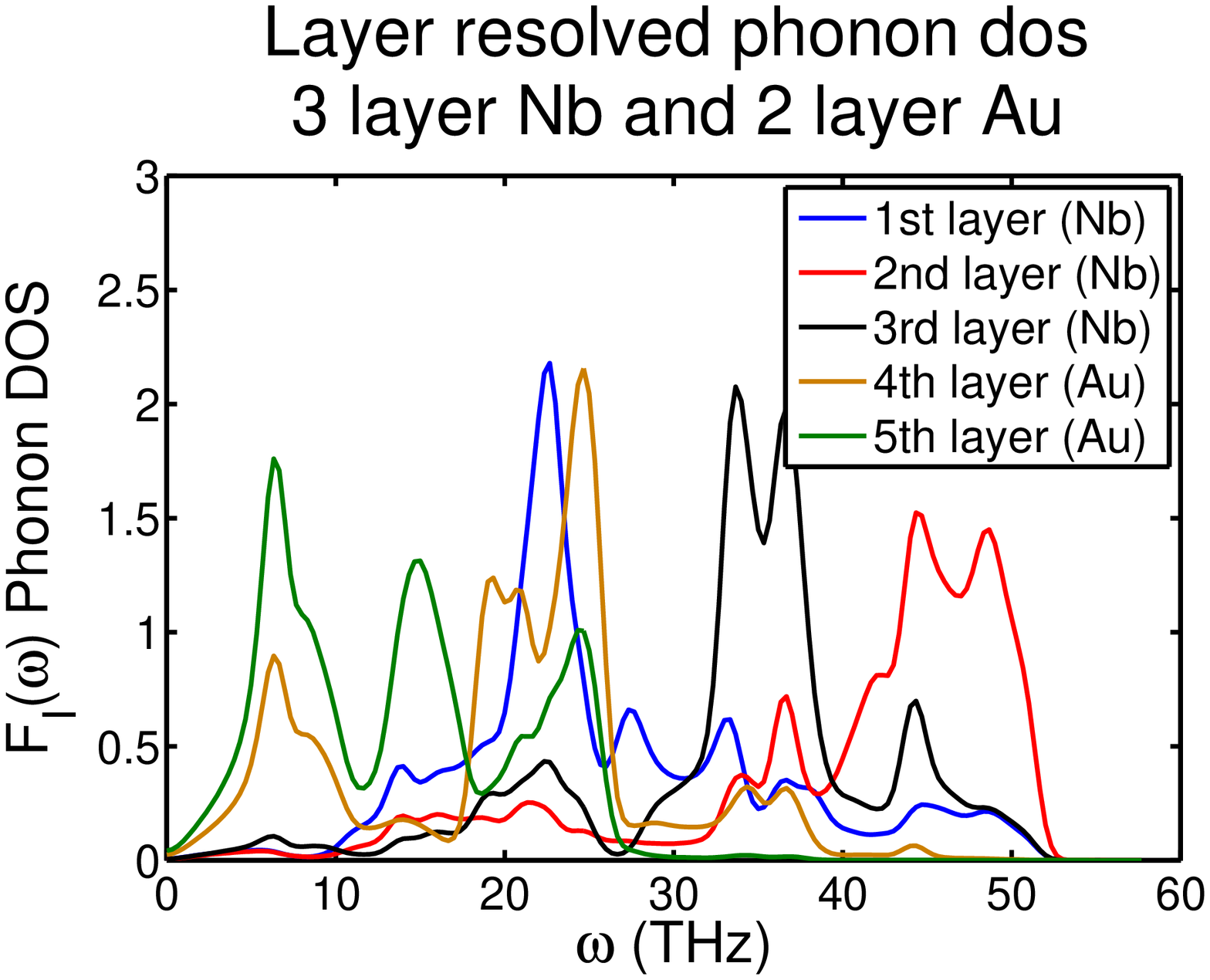}
   \caption{\label{fig:nbau_ph}%
            (Color online) Layer resolved phonon DOS for the following heterostuctures:
            1 Au layer (left panel) and 2 Au layers (right panel) on 3 Nb layers. }
\end{figure}

The more important and more studied systems are the superconducting thin film heterostructures. 
Due to the scarcity of experimental studies of such systems, we choose to investigate the Nb/Au heterostructure, 
mostly because the thick film version was investigated in Refs. \cite{Yamazaki1,Yamazaki2,tcpaper}.
Here 1,2 and 3 layers of gold were added to 3 layers of Nb. We have assumed bcc epitaxial growth for the gold overlayers,
thus the bcc(100) lattice structure is investigated.
The layer resolved phonon DOS is shown for one and two gold overlayers in Fig.~\ref{fig:nbau_ph}.
It can be seen that in the case of a single gold overlayer the phonon spectrum is dominated by low frequencies both in the case of the Au overlayer and 
the top niobium layer (which is on the other side of the slab), therefore, 
the $\left< \omega_I^2 \right>$ becomes smaller on these layers. 
This effect tends to increase the electron-phonon coupling.
However, the McMillan-Hopfield parameter is also smaller for the gold layers as it can be seen in Fig.~\ref{fig:sum_nb} (bottom panel),
since the electronic DOS at the Fermi-level is smaller and 
also the mass of a gold atom is almost twice as large as the mass of a niobium atom. Together these later factors would act to reduce the 
electron-phonon coupling in the gold layers.
% These factors are not present or not as pronounced when the gold coverage increases to two layers or further. 
However, for the 3 Nb/1 Au layers heterostructure, the electron-phonon coupling in the Nb surface layer
is much larger than the one for all other presently investigated heterostructures or slabs.
The net result is an increased overall electron-phonon coupling and (as we will see further down) an increased $T_c$ in the
case of single Au covered Nb thin film.
The results for the different heterostructures are summarized in Fig.~\ref{fig:sum_nb} (bottom panel).

Again, knowing the electron-phonon interaction parameters, 
similar calculations were performed as in the case of the niobium slab to obtain the
critical temperature, the effective electron-phonon coupling, and the effective Coulomb repulsion
as a function of the thickness of the gold overlayers.
In Fig.~\ref{fig:mcm_nb} (right panel) we can observe the inverse proximity effect similarly as it was found
in the Pb/Ag heterostructure in Ref.~\cite{Bourgeois} or in a similar Nb/Au/Nb junction in Ref.~\cite{Yamazaki_arxiv}.
The superconducting transition temperature $T_c$ increases by adding
only one gold overlayer to the niobium, however, adding two layers of gold
does decrease the $T_c$. This result is now well understood based on the previous result regarding the electron-phonon interaction. 
Bourgeois et al.~\cite{Bourgeois} suggested that there is a competition between the Coulomb effects
and the classical proximity effect.
Indeed, in Fig.~\ref{fig:mcm_nb} (right panel) it can be seen that with increasing the number of the gold overlayers the
effective Coulomb repulsion decreases, which can cause an increase in the critical temperature.
Nevertheless, based on  Fig.~\ref{fig:mcm_nb} (right panel) we would rather conclude, that 
the main effect which creates the inverse proximity effect is due to the enhanced electron-phonon coupling in the overlayer. 
The behavior of the electron-phonon interaction appears to primarily influence $T_c$ in other thicknesses as well, overriding the changes 
coming from the Coulomb repulsion. 

In this paper a first-principles approach was presented to investigate superconducting slabs,
and S/N heterostructures.
In essence, the scheme of calculation presented here requires the solution of two separate problems:
solving the KSBdG equations and constructing exchange functionals.
In Ref.~\cite{BdGKKR} the SKKR method was generalized for the superconducting state and
now a simple scheme was constructed to obtain a simple approximation for the exchange functional.
The method was applied to niobium and niobium--gold slabs.
In the case of free standing Nb bcc(100) slabs we have found that the McMillan-Hopfield parameter is larger,
and the $ \left< \omega_I^2 \right>$ frequency is smaller on the surface of the Nb,
which results in large electron-phonon coupling for the surface. As a consequence, the critical temperature is above the bulk value.
For the Nb(110) slab the McMillan-Hopfield parameters are almost constant, and the $ \left< \omega_I^2 \right>$ frequencies
show a behavior similar to that of the Nb(100) surface facet.
Therefore, the critical temperature is oscillating around the bulk value.
While presently there is no first-principles way to calculate the effective Coulomb repulsion parameter ($\mu^*$) directly,
a procedure was developed to estimate this parameter via the McMillan formula.
We also studied the properties of thin Nb/Au heterostructures where we could observe the inverse proximity effect
for which a first-principles based explanation was found.

%% ----------------------------------------------------------------------------
%%    ACKNOWLEDGEMENT
%%
{\emph{Acknowledgment}} ---  Financial support by the Hungarian National, Research, Development and Innovation
Office under the contract No. K115632, the Swedish Research Council, the Swedish Foundation for Strategic Research, and the
Swedish Foundation for International Cooperation in Research and Higher Education is gratefully acknowledged.
The Swedish National Infrastructure for Computing is acknowledged for providing computational facilities.
%% ----------------------------------------------------------------------------
%%
%%    BIBLIOGRAPHY
%%
\bibliographystyle{apsrev4-1}
\bibliography{phonon}

\begin{thebibliography}{34}%
\makeatletter
\providecommand \@ifxundefined [1]{%
 \@ifx{#1\undefined}
}%
\providecommand \@ifnum [1]{%
 \ifnum #1\expandafter \@firstoftwo
 \else \expandafter \@secondoftwo
 \fi
}%
\providecommand \@ifx [1]{%
 \ifx #1\expandafter \@firstoftwo
 \else \expandafter \@secondoftwo
 \fi
}%
\providecommand \natexlab [1]{#1}%
\providecommand \enquote  [1]{``#1''}%
\providecommand \bibnamefont  [1]{#1}%
\providecommand \bibfnamefont [1]{#1}%
\providecommand \citenamefont [1]{#1}%
\providecommand \href@noop [0]{\@secondoftwo}%
\providecommand \href [0]{\begingroup \@sanitize@url \@href}%
\providecommand \@href[1]{\@@startlink{#1}\@@href}%
\providecommand \@@href[1]{\endgroup#1\@@endlink}%
\providecommand \@sanitize@url [0]{\catcode `\\12\catcode `\$12\catcode
  `\&12\catcode `\#12\catcode `\^12\catcode `\_12\catcode `\%12\relax}%
\providecommand \@@startlink[1]{}%
\providecommand \@@endlink[0]{}%
\providecommand \url  [0]{\begingroup\@sanitize@url \@url }%
\providecommand \@url [1]{\endgroup\@href {#1}{\urlprefix }}%
\providecommand \urlprefix  [0]{URL }%
\providecommand \Eprint [0]{\href }%
\providecommand \doibase [0]{http://dx.doi.org/}%
\providecommand \selectlanguage [0]{\@gobble}%
\providecommand \bibinfo  [0]{\@secondoftwo}%
\providecommand \bibfield  [0]{\@secondoftwo}%
\providecommand \translation [1]{[#1]}%
\providecommand \BibitemOpen [0]{}%
\providecommand \bibitemStop [0]{}%
\providecommand \bibitemNoStop [0]{.\EOS\space}%
\providecommand \EOS [0]{\spacefactor3000\relax}%
\providecommand \BibitemShut  [1]{\csname bibitem#1\endcsname}%
\let\auto@bib@innerbib\@empty
%</preamble>
\bibitem [{\citenamefont {Buckel}\ and\ \citenamefont
  {Hilsch}(1954)}]{Buckel1954}%
  \BibitemOpen
  \bibfield  {author} {\bibinfo {author} {\bibfnamefont {W.}~\bibnamefont
  {Buckel}}\ and\ \bibinfo {author} {\bibfnamefont {R.}~\bibnamefont
  {Hilsch}},\ }\href {\doibase 10.1007/bf01337903} {\bibfield  {journal}
  {\bibinfo  {journal} {Zeitschrift f\"ur Physik}\ }\textbf {\bibinfo {volume}
  {138}},\ \bibinfo {pages} {109} (\bibinfo {year} {1954})}\BibitemShut
  {NoStop}%
\bibitem [{\citenamefont {Blatt}\ and\ \citenamefont
  {Thompson}(1963)}]{Blatt1963}%
  \BibitemOpen
  \bibfield  {author} {\bibinfo {author} {\bibfnamefont {J.~M.}\ \bibnamefont
  {Blatt}}\ and\ \bibinfo {author} {\bibfnamefont {C.~J.}\ \bibnamefont
  {Thompson}},\ }\href {\doibase 10.1103/physrevlett.10.332} {\bibfield
  {journal} {\bibinfo  {journal} {Phys. Rev. Lett.}\ }\textbf {\bibinfo
  {volume} {10}},\ \bibinfo {pages} {332} (\bibinfo {year} {1963})}\BibitemShut
  {NoStop}%
\bibitem [{\citenamefont {Abeles}\ \emph {et~al.}(1966)\citenamefont {Abeles},
  \citenamefont {Cohen},\ and\ \citenamefont {Cullen}}]{Abeles1966}%
  \BibitemOpen
  \bibfield  {author} {\bibinfo {author} {\bibfnamefont {B.}~\bibnamefont
  {Abeles}}, \bibinfo {author} {\bibfnamefont {R.~W.}\ \bibnamefont {Cohen}}, \
  and\ \bibinfo {author} {\bibfnamefont {G.~W.}\ \bibnamefont {Cullen}},\
  }\href {\doibase 10.1103/physrevlett.17.632} {\bibfield  {journal} {\bibinfo
  {journal} {Phys. Rev. Lett.}\ }\textbf {\bibinfo {volume} {17}},\ \bibinfo
  {pages} {632} (\bibinfo {year} {1966})}\BibitemShut {NoStop}%
\bibitem [{\citenamefont {Strongin}\ \emph {et~al.}(1970)\citenamefont
  {Strongin}, \citenamefont {Thompson}, \citenamefont {Kammerer},\ and\
  \citenamefont {Crow}}]{Strongin}%
  \BibitemOpen
  \bibfield  {author} {\bibinfo {author} {\bibfnamefont {M.}~\bibnamefont
  {Strongin}}, \bibinfo {author} {\bibfnamefont {R.~S.}\ \bibnamefont
  {Thompson}}, \bibinfo {author} {\bibfnamefont {O.~F.}\ \bibnamefont
  {Kammerer}}, \ and\ \bibinfo {author} {\bibfnamefont {J.~E.}\ \bibnamefont
  {Crow}},\ }\href {\doibase 10.1103/PhysRevB.1.1078} {\bibfield  {journal}
  {\bibinfo  {journal} {Phys. Rev. B}\ }\textbf {\bibinfo {volume} {1}},\
  \bibinfo {pages} {1078} (\bibinfo {year} {1970})}\BibitemShut {NoStop}%
\bibitem [{\citenamefont {Haviland}\ \emph {et~al.}(1989)\citenamefont
  {Haviland}, \citenamefont {Liu},\ and\ \citenamefont
  {Goldman}}]{Haviland1989}%
  \BibitemOpen
  \bibfield  {author} {\bibinfo {author} {\bibfnamefont {D.~B.}\ \bibnamefont
  {Haviland}}, \bibinfo {author} {\bibfnamefont {Y.}~\bibnamefont {Liu}}, \
  and\ \bibinfo {author} {\bibfnamefont {A.~M.}\ \bibnamefont {Goldman}},\
  }\href {\doibase 10.1103/physrevlett.62.2180} {\bibfield  {journal} {\bibinfo
   {journal} {Phys. Rev. Lett.}\ }\textbf {\bibinfo {volume} {62}},\ \bibinfo
  {pages} {2180} (\bibinfo {year} {1989})}\BibitemShut {NoStop}%
\bibitem [{\citenamefont {Bourgeois}\ \emph {et~al.}(2002)\citenamefont
  {Bourgeois}, \citenamefont {Frydman},\ and\ \citenamefont
  {Dynes}}]{Bourgeois}%
  \BibitemOpen
  \bibfield  {author} {\bibinfo {author} {\bibfnamefont {O.}~\bibnamefont
  {Bourgeois}}, \bibinfo {author} {\bibfnamefont {A.}~\bibnamefont {Frydman}},
  \ and\ \bibinfo {author} {\bibfnamefont {R.~C.}\ \bibnamefont {Dynes}},\
  }\href {\doibase 10.1103/PhysRevLett.88.186403} {\bibfield  {journal}
  {\bibinfo  {journal} {Phys. Rev. Lett.}\ }\textbf {\bibinfo {volume} {88}},\
  \bibinfo {pages} {186403} (\bibinfo {year} {2002})}\BibitemShut {NoStop}%
\bibitem [{\citenamefont {Guo}(2004)}]{Guo2004}%
  \BibitemOpen
  \bibfield  {author} {\bibinfo {author} {\bibfnamefont {Y.}~\bibnamefont
  {Guo}},\ }\href {\doibase 10.1126/science.1105130} {\bibfield  {journal}
  {\bibinfo  {journal} {Science}\ }\textbf {\bibinfo {volume} {306}},\ \bibinfo
  {pages} {1915} (\bibinfo {year} {2004})}\BibitemShut {NoStop}%
\bibitem [{\citenamefont {\"{O}zer}\ \emph {et~al.}(2006)\citenamefont
  {\"{O}zer}, \citenamefont {Thompson},\ and\ \citenamefont
  {Weitering}}]{zer2006}%
  \BibitemOpen
  \bibfield  {author} {\bibinfo {author} {\bibfnamefont {M.~M.}\ \bibnamefont
  {\"{O}zer}}, \bibinfo {author} {\bibfnamefont {J.~R.}\ \bibnamefont
  {Thompson}}, \ and\ \bibinfo {author} {\bibfnamefont {H.~H.}\ \bibnamefont
  {Weitering}},\ }\href {\doibase 10.1038/nphys244} {\bibfield  {journal}
  {\bibinfo  {journal} {Nat Phys}\ }\textbf {\bibinfo {volume} {2}},\ \bibinfo
  {pages} {173} (\bibinfo {year} {2006})}\BibitemShut {NoStop}%
\bibitem [{\citenamefont {Eom}\ \emph {et~al.}(2006)\citenamefont {Eom},
  \citenamefont {Qin}, \citenamefont {Chou},\ and\ \citenamefont
  {Shih}}]{Eom2006}%
  \BibitemOpen
  \bibfield  {author} {\bibinfo {author} {\bibfnamefont {D.}~\bibnamefont
  {Eom}}, \bibinfo {author} {\bibfnamefont {S.}~\bibnamefont {Qin}}, \bibinfo
  {author} {\bibfnamefont {M.-Y.}\ \bibnamefont {Chou}}, \ and\ \bibinfo
  {author} {\bibfnamefont {C.~K.}\ \bibnamefont {Shih}},\ }\href {\doibase
  10.1103/physrevlett.96.027005} {\bibfield  {journal} {\bibinfo  {journal}
  {Phys. Rev. Lett.}\ }\textbf {\bibinfo {volume} {96}} (\bibinfo {year}
  {2006}),\ 10.1103/physrevlett.96.027005}\BibitemShut {NoStop}%
\bibitem [{\citenamefont {Brun}\ \emph {et~al.}(2009)\citenamefont {Brun},
  \citenamefont {Hong}, \citenamefont {Patthey}, \citenamefont {Sklyadneva},
  \citenamefont {Heid}, \citenamefont {Echenique}, \citenamefont {Bohnen},
  \citenamefont {Chulkov},\ and\ \citenamefont {Schneider}}]{Brun2009}%
  \BibitemOpen
  \bibfield  {author} {\bibinfo {author} {\bibfnamefont {C.}~\bibnamefont
  {Brun}}, \bibinfo {author} {\bibfnamefont {I.-P.}\ \bibnamefont {Hong}},
  \bibinfo {author} {\bibfnamefont {F.}~\bibnamefont {Patthey}}, \bibinfo
  {author} {\bibfnamefont {I.~Y.}\ \bibnamefont {Sklyadneva}}, \bibinfo
  {author} {\bibfnamefont {R.}~\bibnamefont {Heid}}, \bibinfo {author}
  {\bibfnamefont {P.~M.}\ \bibnamefont {Echenique}}, \bibinfo {author}
  {\bibfnamefont {K.~P.}\ \bibnamefont {Bohnen}}, \bibinfo {author}
  {\bibfnamefont {E.~V.}\ \bibnamefont {Chulkov}}, \ and\ \bibinfo {author}
  {\bibfnamefont {W.-D.}\ \bibnamefont {Schneider}},\ }\href {\doibase
  10.1103/physrevlett.102.207002} {\bibfield  {journal} {\bibinfo  {journal}
  {Phys. Rev. Lett.}\ }\textbf {\bibinfo {volume} {102}} (\bibinfo {year}
  {2009}),\ 10.1103/physrevlett.102.207002}\BibitemShut {NoStop}%
\bibitem [{\citenamefont {Wang}\ \emph {et~al.}(2009)\citenamefont {Wang},
  \citenamefont {Zhang}, \citenamefont {Loy}, \citenamefont {Chiang},\ and\
  \citenamefont {Xiao}}]{Wang2009}%
  \BibitemOpen
  \bibfield  {author} {\bibinfo {author} {\bibfnamefont {K.}~\bibnamefont
  {Wang}}, \bibinfo {author} {\bibfnamefont {X.}~\bibnamefont {Zhang}},
  \bibinfo {author} {\bibfnamefont {M.~M.~T.}\ \bibnamefont {Loy}}, \bibinfo
  {author} {\bibfnamefont {T.-C.}\ \bibnamefont {Chiang}}, \ and\ \bibinfo
  {author} {\bibfnamefont {X.}~\bibnamefont {Xiao}},\ }\href {\doibase
  10.1103/physrevlett.102.076801} {\bibfield  {journal} {\bibinfo  {journal}
  {Phys. Rev. Lett.}\ }\textbf {\bibinfo {volume} {102}} (\bibinfo {year}
  {2009}),\ 10.1103/physrevlett.102.076801}\BibitemShut {NoStop}%
\bibitem [{\citenamefont {Qin}\ \emph {et~al.}(2009)\citenamefont {Qin},
  \citenamefont {Kim}, \citenamefont {Niu},\ and\ \citenamefont
  {Shih}}]{Qin2009}%
  \BibitemOpen
  \bibfield  {author} {\bibinfo {author} {\bibfnamefont {S.}~\bibnamefont
  {Qin}}, \bibinfo {author} {\bibfnamefont {J.}~\bibnamefont {Kim}}, \bibinfo
  {author} {\bibfnamefont {Q.}~\bibnamefont {Niu}}, \ and\ \bibinfo {author}
  {\bibfnamefont {C.-K.}\ \bibnamefont {Shih}},\ }\href {\doibase
  10.1126/science.1170775} {\bibfield  {journal} {\bibinfo  {journal}
  {Science}\ }\textbf {\bibinfo {volume} {324}},\ \bibinfo {pages} {1314}
  (\bibinfo {year} {2009})}\BibitemShut {NoStop}%
\bibitem [{\citenamefont {Zhang}\ \emph {et~al.}(2010)\citenamefont {Zhang},
  \citenamefont {Cheng}, \citenamefont {Li}, \citenamefont {Sun}, \citenamefont
  {Wang}, \citenamefont {Zhu}, \citenamefont {He}, \citenamefont {Wang},
  \citenamefont {Ma}, \citenamefont {Chen}, \citenamefont {Wang}, \citenamefont
  {Liu}, \citenamefont {Lin}, \citenamefont {Jia},\ and\ \citenamefont
  {Xue}}]{Zhang}%
  \BibitemOpen
  \bibfield  {author} {\bibinfo {author} {\bibfnamefont {T.}~\bibnamefont
  {Zhang}}, \bibinfo {author} {\bibfnamefont {P.}~\bibnamefont {Cheng}},
  \bibinfo {author} {\bibfnamefont {W.-J.}\ \bibnamefont {Li}}, \bibinfo
  {author} {\bibfnamefont {Y.-J.}\ \bibnamefont {Sun}}, \bibinfo {author}
  {\bibfnamefont {G.}~\bibnamefont {Wang}}, \bibinfo {author} {\bibfnamefont
  {X.-G.}\ \bibnamefont {Zhu}}, \bibinfo {author} {\bibfnamefont
  {K.}~\bibnamefont {He}}, \bibinfo {author} {\bibfnamefont {L.}~\bibnamefont
  {Wang}}, \bibinfo {author} {\bibfnamefont {X.}~\bibnamefont {Ma}}, \bibinfo
  {author} {\bibfnamefont {X.}~\bibnamefont {Chen}}, \bibinfo {author}
  {\bibfnamefont {Y.}~\bibnamefont {Wang}}, \bibinfo {author} {\bibfnamefont
  {Y.}~\bibnamefont {Liu}}, \bibinfo {author} {\bibfnamefont {H.-Q.}\
  \bibnamefont {Lin}}, \bibinfo {author} {\bibfnamefont {J.-F.}\ \bibnamefont
  {Jia}}, \ and\ \bibinfo {author} {\bibfnamefont {Q.-K.}\ \bibnamefont
  {Xue}},\ }\href {\doibase 10.1038/nphys1499} {\bibfield  {journal} {\bibinfo
  {journal} {Nat Phys}\ }\textbf {\bibinfo {volume} {6}},\ \bibinfo {pages}
  {104} (\bibinfo {year} {2010})}\BibitemShut {NoStop}%
\bibitem [{\citenamefont {Yamazaki}\ \emph {et~al.}(2006)\citenamefont
  {Yamazaki}, \citenamefont {Shannon},\ and\ \citenamefont
  {Takagi}}]{Yamazaki1}%
  \BibitemOpen
  \bibfield  {author} {\bibinfo {author} {\bibfnamefont {H.}~\bibnamefont
  {Yamazaki}}, \bibinfo {author} {\bibfnamefont {N.}~\bibnamefont {Shannon}}, \
  and\ \bibinfo {author} {\bibfnamefont {H.}~\bibnamefont {Takagi}},\ }\href
  {\doibase 10.1103/PhysRevB.73.094507} {\bibfield  {journal} {\bibinfo
  {journal} {Phys. Rev. B}\ }\textbf {\bibinfo {volume} {73}},\ \bibinfo
  {pages} {094507} (\bibinfo {year} {2006})}\BibitemShut {NoStop}%
\bibitem [{\citenamefont {Yamazaki}\ \emph {et~al.}(2010)\citenamefont
  {Yamazaki}, \citenamefont {Shannon},\ and\ \citenamefont
  {Takagi}}]{Yamazaki2}%
  \BibitemOpen
  \bibfield  {author} {\bibinfo {author} {\bibfnamefont {H.}~\bibnamefont
  {Yamazaki}}, \bibinfo {author} {\bibfnamefont {N.}~\bibnamefont {Shannon}}, \
  and\ \bibinfo {author} {\bibfnamefont {H.}~\bibnamefont {Takagi}},\ }\href
  {\doibase 10.1103/PhysRevB.81.094503} {\bibfield  {journal} {\bibinfo
  {journal} {Phys. Rev. B}\ }\textbf {\bibinfo {volume} {81}},\ \bibinfo
  {pages} {094503} (\bibinfo {year} {2010})}\BibitemShut {NoStop}%
\bibitem [{\citenamefont {Csire}\ \emph {et~al.}(2016)\citenamefont {Csire},
  \citenamefont {Cserti}, \citenamefont {T{\" u}tt{\H o}},\ and\ \citenamefont
  {\'Ujfalussy}}]{tcpaper}%
  \BibitemOpen
  \bibfield  {author} {\bibinfo {author} {\bibfnamefont {G.}~\bibnamefont
  {Csire}}, \bibinfo {author} {\bibfnamefont {J.}~\bibnamefont {Cserti}},
  \bibinfo {author} {\bibfnamefont {I.}~\bibnamefont {T{\" u}tt{\H o}}}, \ and\
  \bibinfo {author} {\bibfnamefont {B.}~\bibnamefont {\'Ujfalussy}},\ }\href
  {http://arxiv.org/abs/1601.07038} {\bibfield  {journal} {\bibinfo  {journal}
  {arXiv:1601.07038 [cond-mat.supr-con]}\ } (\bibinfo {year}
  {2016})}\BibitemShut {NoStop}%
\bibitem [{\citenamefont {Oliveira}\ \emph {et~al.}(1988)\citenamefont
  {Oliveira}, \citenamefont {Gross},\ and\ \citenamefont {Kohn}}]{OGK}%
  \BibitemOpen
  \bibfield  {author} {\bibinfo {author} {\bibfnamefont {L.~N.}\ \bibnamefont
  {Oliveira}}, \bibinfo {author} {\bibfnamefont {E.~K.~U.}\ \bibnamefont
  {Gross}}, \ and\ \bibinfo {author} {\bibfnamefont {W.}~\bibnamefont {Kohn}},\
  }\href {\doibase 10.1103/PhysRevLett.60.2430} {\bibfield  {journal} {\bibinfo
   {journal} {Phys. Rev. Lett.}\ }\textbf {\bibinfo {volume} {60}},\ \bibinfo
  {pages} {2430} (\bibinfo {year} {1988})}\BibitemShut {NoStop}%
\bibitem [{\citenamefont {L\"uders}\ \emph {et~al.}(2005)\citenamefont
  {L\"uders}, \citenamefont {Marques}, \citenamefont {Lathiotakis},
  \citenamefont {Floris}, \citenamefont {Profeta}, \citenamefont {Fast},
  \citenamefont {Continenza}, \citenamefont {Massidda},\ and\ \citenamefont
  {Gross}}]{Luders1}%
  \BibitemOpen
  \bibfield  {author} {\bibinfo {author} {\bibfnamefont {M.}~\bibnamefont
  {L\"uders}}, \bibinfo {author} {\bibfnamefont {M.~A.~L.}\ \bibnamefont
  {Marques}}, \bibinfo {author} {\bibfnamefont {N.~N.}\ \bibnamefont
  {Lathiotakis}}, \bibinfo {author} {\bibfnamefont {A.}~\bibnamefont {Floris}},
  \bibinfo {author} {\bibfnamefont {G.}~\bibnamefont {Profeta}}, \bibinfo
  {author} {\bibfnamefont {L.}~\bibnamefont {Fast}}, \bibinfo {author}
  {\bibfnamefont {A.}~\bibnamefont {Continenza}}, \bibinfo {author}
  {\bibfnamefont {S.}~\bibnamefont {Massidda}}, \ and\ \bibinfo {author}
  {\bibfnamefont {E.~K.~U.}\ \bibnamefont {Gross}},\ }\href {\doibase
  10.1103/PhysRevB.72.024545} {\bibfield  {journal} {\bibinfo  {journal} {Phys.
  Rev. B}\ }\textbf {\bibinfo {volume} {72}},\ \bibinfo {pages} {024545}
  (\bibinfo {year} {2005})}\BibitemShut {NoStop}%
\bibitem [{\citenamefont {Marques}\ \emph {et~al.}(2005)\citenamefont
  {Marques}, \citenamefont {L\"uders}, \citenamefont {Lathiotakis},
  \citenamefont {Profeta}, \citenamefont {Floris}, \citenamefont {Fast},
  \citenamefont {Continenza}, \citenamefont {Gross},\ and\ \citenamefont
  {Massidda}}]{Luders2}%
  \BibitemOpen
  \bibfield  {author} {\bibinfo {author} {\bibfnamefont {M.~A.~L.}\
  \bibnamefont {Marques}}, \bibinfo {author} {\bibfnamefont {M.}~\bibnamefont
  {L\"uders}}, \bibinfo {author} {\bibfnamefont {N.~N.}\ \bibnamefont
  {Lathiotakis}}, \bibinfo {author} {\bibfnamefont {G.}~\bibnamefont
  {Profeta}}, \bibinfo {author} {\bibfnamefont {A.}~\bibnamefont {Floris}},
  \bibinfo {author} {\bibfnamefont {L.}~\bibnamefont {Fast}}, \bibinfo {author}
  {\bibfnamefont {A.}~\bibnamefont {Continenza}}, \bibinfo {author}
  {\bibfnamefont {E.~K.~U.}\ \bibnamefont {Gross}}, \ and\ \bibinfo {author}
  {\bibfnamefont {S.}~\bibnamefont {Massidda}},\ }\href {\doibase
  10.1103/PhysRevB.72.024546} {\bibfield  {journal} {\bibinfo  {journal} {Phys.
  Rev. B}\ }\textbf {\bibinfo {volume} {72}},\ \bibinfo {pages} {024546}
  (\bibinfo {year} {2005})}\BibitemShut {NoStop}%
\bibitem [{\citenamefont {Csire}\ \emph {et~al.}(2015)\citenamefont {Csire},
  \citenamefont {\'Ujfalussy}, \citenamefont {Cserti},\ and\ \citenamefont
  {Gy\ifmmode~\mbox{\H{o}}\else \H{o}\fi{}rffy}}]{BdGKKR}%
  \BibitemOpen
  \bibfield  {author} {\bibinfo {author} {\bibfnamefont {G.}~\bibnamefont
  {Csire}}, \bibinfo {author} {\bibfnamefont {B.}~\bibnamefont {\'Ujfalussy}},
  \bibinfo {author} {\bibfnamefont {J.}~\bibnamefont {Cserti}}, \ and\ \bibinfo
  {author} {\bibfnamefont {B.}~\bibnamefont {Gy\ifmmode~\mbox{\H{o}}\else
  \H{o}\fi{}rffy}},\ }\href {\doibase 10.1103/PhysRevB.91.165142} {\bibfield
  {journal} {\bibinfo  {journal} {Phys. Rev. B}\ }\textbf {\bibinfo {volume}
  {91}},\ \bibinfo {pages} {165142} (\bibinfo {year} {2015})}\BibitemShut
  {NoStop}%
\bibitem [{\citenamefont {Suvasini}\ \emph {et~al.}(1993)\citenamefont
  {Suvasini}, \citenamefont {Temmerman},\ and\ \citenamefont
  {Gyorffy}}]{Suvasini}%
  \BibitemOpen
  \bibfield  {author} {\bibinfo {author} {\bibfnamefont {M.~B.}\ \bibnamefont
  {Suvasini}}, \bibinfo {author} {\bibfnamefont {W.~M.}\ \bibnamefont
  {Temmerman}}, \ and\ \bibinfo {author} {\bibfnamefont {B.~L.}\ \bibnamefont
  {Gyorffy}},\ }\href {\doibase 10.1103/PhysRevB.48.1202} {\bibfield  {journal}
  {\bibinfo  {journal} {Phys. Rev. B}\ }\textbf {\bibinfo {volume} {48}},\
  \bibinfo {pages} {1202} (\bibinfo {year} {1993})}\BibitemShut {NoStop}%
\bibitem [{\citenamefont {McMillan}(1968)}]{McMillan}%
  \BibitemOpen
  \bibfield  {author} {\bibinfo {author} {\bibfnamefont {W.~L.}\ \bibnamefont
  {McMillan}},\ }\href {\doibase 10.1103/PhysRev.167.331} {\bibfield  {journal}
  {\bibinfo  {journal} {Phys. Rev.}\ }\textbf {\bibinfo {volume} {167}},\
  \bibinfo {pages} {331} (\bibinfo {year} {1968})}\BibitemShut {NoStop}%
\bibitem [{\citenamefont {Gaspari}\ and\ \citenamefont
  {Gyorffy}(1972)}]{Gaspari}%
  \BibitemOpen
  \bibfield  {author} {\bibinfo {author} {\bibfnamefont {G.~D.}\ \bibnamefont
  {Gaspari}}\ and\ \bibinfo {author} {\bibfnamefont {B.~L.}\ \bibnamefont
  {Gyorffy}},\ }\href {\doibase 10.1103/PhysRevLett.28.801} {\bibfield
  {journal} {\bibinfo  {journal} {Phys. Rev. Lett.}\ }\textbf {\bibinfo
  {volume} {28}},\ \bibinfo {pages} {801} (\bibinfo {year} {1972})}\BibitemShut
  {NoStop}%
\bibitem [{\citenamefont {Savrasov}\ \emph {et~al.}(1994)\citenamefont
  {Savrasov}, \citenamefont {Savrasov},\ and\ \citenamefont
  {Andersen}}]{Savrasov}%
  \BibitemOpen
  \bibfield  {author} {\bibinfo {author} {\bibfnamefont {S.~Y.}\ \bibnamefont
  {Savrasov}}, \bibinfo {author} {\bibfnamefont {D.~Y.}\ \bibnamefont
  {Savrasov}}, \ and\ \bibinfo {author} {\bibfnamefont {O.~K.}\ \bibnamefont
  {Andersen}},\ }\href {\doibase 10.1103/PhysRevLett.72.372} {\bibfield
  {journal} {\bibinfo  {journal} {Phys. Rev. Lett.}\ }\textbf {\bibinfo
  {volume} {72}},\ \bibinfo {pages} {372} (\bibinfo {year} {1994})}\BibitemShut
  {NoStop}%
\bibitem [{\citenamefont {Savrasov}(1996)}]{Savrasov2}%
  \BibitemOpen
  \bibfield  {author} {\bibinfo {author} {\bibfnamefont {S.~Y.}\ \bibnamefont
  {Savrasov}},\ }\href {\doibase 10.1103/PhysRevB.54.16470} {\bibfield
  {journal} {\bibinfo  {journal} {Phys. Rev. B}\ }\textbf {\bibinfo {volume}
  {54}},\ \bibinfo {pages} {16470} (\bibinfo {year} {1996})}\BibitemShut
  {NoStop}%
\bibitem [{\citenamefont {Baroni}\ \emph {et~al.}(2001)\citenamefont {Baroni},
  \citenamefont {de~Gironcoli}, \citenamefont {Corso},\ and\ \citenamefont
  {Giannozzi}}]{Baroni2001}%
  \BibitemOpen
  \bibfield  {author} {\bibinfo {author} {\bibfnamefont {S.}~\bibnamefont
  {Baroni}}, \bibinfo {author} {\bibfnamefont {S.}~\bibnamefont
  {de~Gironcoli}}, \bibinfo {author} {\bibfnamefont {A.~D.}\ \bibnamefont
  {Corso}}, \ and\ \bibinfo {author} {\bibfnamefont {P.}~\bibnamefont
  {Giannozzi}},\ }\href {\doibase 10.1103/revmodphys.73.515} {\bibfield
  {journal} {\bibinfo  {journal} {Reviews of Modern Physics}\ }\textbf
  {\bibinfo {volume} {73}},\ \bibinfo {pages} {515} (\bibinfo {year}
  {2001})}\BibitemShut {NoStop}%
\bibitem [{\citenamefont {Kresse}\ and\ \citenamefont
  {Furthm\"{u}ller}(1996)}]{Kresse1996}%
  \BibitemOpen
  \bibfield  {author} {\bibinfo {author} {\bibfnamefont {G.}~\bibnamefont
  {Kresse}}\ and\ \bibinfo {author} {\bibfnamefont {J.}~\bibnamefont
  {Furthm\"{u}ller}},\ }\href {\doibase 10.1103/physrevb.54.11169} {\bibfield
  {journal} {\bibinfo  {journal} {Phys. Rev. B}\ }\textbf {\bibinfo {volume}
  {54}},\ \bibinfo {pages} {11169} (\bibinfo {year} {1996})}\BibitemShut
  {NoStop}%
\bibitem [{\citenamefont {Togo}\ and\ \citenamefont {Tanaka}(2015)}]{Togo2015}%
  \BibitemOpen
  \bibfield  {author} {\bibinfo {author} {\bibfnamefont {A.}~\bibnamefont
  {Togo}}\ and\ \bibinfo {author} {\bibfnamefont {I.}~\bibnamefont {Tanaka}},\
  }\href {\doibase 10.1016/j.scriptamat.2015.07.021} {\bibfield  {journal}
  {\bibinfo  {journal} {Scripta Materialia}\ }\textbf {\bibinfo {volume}
  {108}},\ \bibinfo {pages} {1} (\bibinfo {year} {2015})}\BibitemShut {NoStop}%
\bibitem [{\citenamefont {Klein}\ and\ \citenamefont
  {Papaconstantopoulos}(1974)}]{Klein}%
  \BibitemOpen
  \bibfield  {author} {\bibinfo {author} {\bibfnamefont {B.~M.}\ \bibnamefont
  {Klein}}\ and\ \bibinfo {author} {\bibfnamefont {D.~A.}\ \bibnamefont
  {Papaconstantopoulos}},\ }\href {\doibase 10.1103/PhysRevLett.32.1193}
  {\bibfield  {journal} {\bibinfo  {journal} {Phys. Rev. Lett.}\ }\textbf
  {\bibinfo {volume} {32}},\ \bibinfo {pages} {1193} (\bibinfo {year}
  {1974})}\BibitemShut {NoStop}%
\bibitem [{\citenamefont {Allen}\ and\ \citenamefont
  {Mitrovi{\'{c}}}(1983)}]{Allen1983}%
  \BibitemOpen
  \bibfield  {author} {\bibinfo {author} {\bibfnamefont {P.~B.}\ \bibnamefont
  {Allen}}\ and\ \bibinfo {author} {\bibfnamefont {B.}~\bibnamefont
  {Mitrovi{\'{c}}}},\ }in\ \href {\doibase 10.1016/s0081-1947(08)60665-7}
  {\emph {\bibinfo {booktitle} {Solid State Physics}}}\ (\bibinfo  {publisher}
  {Elsevier {BV}},\ \bibinfo {year} {1983})\ pp.\ \bibinfo {pages}
  {1--92}\BibitemShut {NoStop}%
\bibitem [{\citenamefont {Csire}\ \emph {et~al.}()\citenamefont {Csire},
  \citenamefont {Cserti}, \citenamefont {T{\" u}tt{\H o}},\ and\ \citenamefont
  {\'Ujfalussy}}]{Proximity}%
  \BibitemOpen
  \bibfield  {author} {\bibinfo {author} {\bibfnamefont {G.}~\bibnamefont
  {Csire}}, \bibinfo {author} {\bibfnamefont {J.}~\bibnamefont {Cserti}},
  \bibinfo {author} {\bibfnamefont {I.}~\bibnamefont {T{\" u}tt{\H o}}}, \ and\
  \bibinfo {author} {\bibfnamefont {B.}~\bibnamefont {\'Ujfalussy}},\
  }\href@noop {} {\bibinfo  {journal} {unpublished}\ }\BibitemShut {NoStop}%
\bibitem [{\citenamefont {Nicol}\ and\ \citenamefont
  {Carbotte}(2005)}]{Nicol2005}%
  \BibitemOpen
\bibfield  {journal} {  }\bibfield  {author} {\bibinfo {author} {\bibfnamefont
  {E.~J.}\ \bibnamefont {Nicol}}\ and\ \bibinfo {author} {\bibfnamefont
  {J.~P.}\ \bibnamefont {Carbotte}},\ }\href {\doibase
  10.1103/physrevb.71.054501} {\bibfield  {journal} {\bibinfo  {journal} {Phys.
  Rev. B}\ }\textbf {\bibinfo {volume} {71}} (\bibinfo {year} {2005}),\
  10.1103/physrevb.71.054501}\BibitemShut {NoStop}%
\bibitem [{\citenamefont {De~Gennes}(1964)}]{deGennes}%
  \BibitemOpen
  \bibfield  {author} {\bibinfo {author} {\bibfnamefont {P.~G.}\ \bibnamefont
  {De~Gennes}},\ }\href {\doibase 10.1103/RevModPhys.36.225} {\bibfield
  {journal} {\bibinfo  {journal} {Rev. Mod. Phys.}\ }\textbf {\bibinfo {volume}
  {36}},\ \bibinfo {pages} {225} (\bibinfo {year} {1964})}\BibitemShut
  {NoStop}%
\bibitem [{\citenamefont {Yamazaki}\ \emph {et~al.}(2016)\citenamefont
  {Yamazaki}, \citenamefont {Shannon},\ and\ \citenamefont
  {Takagi}}]{Yamazaki_arxiv}%
  \BibitemOpen
  \bibfield  {author} {\bibinfo {author} {\bibfnamefont {H.}~\bibnamefont
  {Yamazaki}}, \bibinfo {author} {\bibfnamefont {N.}~\bibnamefont {Shannon}}, \
  and\ \bibinfo {author} {\bibfnamefont {H.}~\bibnamefont {Takagi}},\ }\href
  {http://arxiv.org/abs/1602.05790} {\bibfield  {journal} {\bibinfo  {journal}
  {arXiv:1602.05790 [cond-mat.supr-con]}\ } (\bibinfo {year}
  {2016})}\BibitemShut {NoStop}%
\end{thebibliography}%

\end{document}